\documentclass[%
superscriptaddress,
frontmatterverbose, 
 amsmath,amssymb,
 aps,
pra,
]{revtex4-2}

\usepackage{graphicx}
\usepackage{dcolumn}
\usepackage{bm}
\usepackage{hyperref}
\usepackage[mathlines]{lineno}
\usepackage{color}
\newcommand{\edit}[1]{{\color{black}#1}}
\newcommand{\newedit}[1]{{\color{black}#1}}

\begin{document}

\preprint{Synchronization of wave-propelled capillary spinners}

\title{Synchronization of wave-propelled capillary spinners}

\author{Jack-William Barotta}
 \email{jack-william\_barotta@brown.edu}
\affiliation{
School of Engineering, Center for Fluid Mechanics,
Brown University, 184 Hope Street, Providence, RI 02912}

\author{Giuseppe Pucci}

 \affiliation{%
Consiglio Nazionale delle Ricerche - Istituto di Nanotecnologia (CNR-Nanotec), Via P. Bucci 33C, 87036 Rende, Italy
}
\affiliation{%
INFN, Sezione di Lecce, Via per Monteroni, Lecce 73100, Italy
}

\author{Eli Silver}
\affiliation{
School of Engineering, Center for Fluid Mechanics,
Brown University, 184 Hope Street, Providence, RI 02912}

\author{Alireza Hooshanginejad}
\affiliation{
School of Engineering, Center for Fluid Mechanics,
Brown University, 184 Hope Street, Providence, RI 02912}

\author{Daniel M. Harris}
 \email{daniel\_harris3@brown.edu}
\affiliation{
School of Engineering, Center for Fluid Mechanics,
Brown University, 184 Hope Street, Providence, RI 02912}

\date{\today}
\maketitle

\section*{Abstract}
\textbf{When a millimetric body is placed atop a vibrating liquid bath, the relative motion between the object and interface generates outward propagating waves with an associated momentum flux. Prior work has shown that isolated chiral objects, referred to as spinners, can thus rotate steadily in response to their self-generated wavefield.  Here, we consider the case of two co-chiral spinners held at a fixed spacing from one another but otherwise free to interact hydrodynamically through their shared fluid substrate.  Two identical spinners are able to synchronize their rotation, with their equilibrium phase difference sensitive to their spacing and initial conditions, and even cease to rotate when the coupling becomes sufficiently strong.  Non-identical spinners can also find synchrony provided their intrinsic differences are not too disparate. A hydrodynamic wave model of the spinner interaction is proposed, recovering all salient features of the experiment. In all cases, the spatially periodic nature of the capillary wave coupling is directly reflected in the emergent equilibrium behaviors.}

\section*{Introduction}

Bedridden from a sickness in 1665, Christiaan Huygens observed the coupling of two pendulum clocks  \cite{huygens1895letters}, noticing that the two clocks would begin to oscillate perfectly out of phase when positioned near one another. Realizing that the two oscillators were able to ``communicate'' via a shared substrate of a heavy beam on which the two clocks were hung, he documented this communication as a ``sympathy'' the oscillators had towards one another. Over 300 years later, Huygens' discovery was shown to be one of both ``talent and luck'': the weight of the clocks was just right, and the frequencies were just so, such that synchronization could occur in his realization \cite{bennett2002huygens}.  Numerous followup studies have continued to interrogate the vast parameter space \cite{oliveira2015huygens, willms2017huygens} and explore general complex phase oscillator networks \cite{martens2013chimera}. An abundance of other artificial and natural systems exhibiting synchronization have since been discovered, such as forests of blinking fireflies \cite{buck1966biology, buck1968mechanism, sarfati2020spatio}, human crowds on bridges \cite{fujino1993synchronization, eckhardt2007modeling}, de-tuned pipe organs \cite{rayleigh1882pitch, abel2009synchronization}, and even the calls of male frogs \cite{aihara2008mathematical}.  However in many cases, companion mathematical models reproducing such phenomena are purely phenomenological due to the inherent complexity or inaccessibility of the systems.  In fact, despite its rich history and apparent simplicity, even the underlying physics of the pendulum system remains an active discussion with much still yet to be learned \cite{goldsztein2022coupled}.

\begin{figure*}
    \centering
    \includegraphics{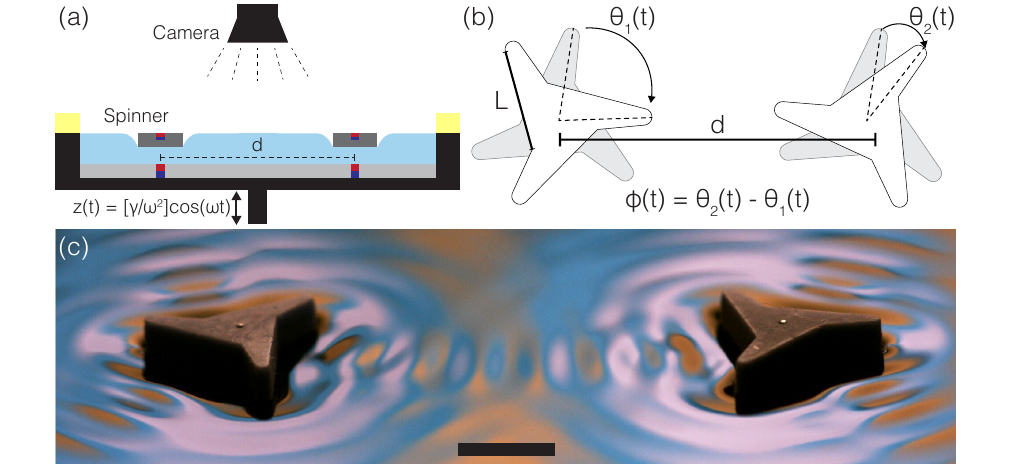}
    \caption{ \textbf{A pair of spinners interact via their mutually generated wavefield.} \textbf{(a)} A schematic of the experimental set-up: two spinners are placed on a vertically vibrating bath at a distance $d$ apart, translationally fixed by magnets in and below the spinners. The bath is oscillated vertically with position $z(t) = \gamma/\omega^2\cos\left(\omega t\right)$, where $\gamma$ is the acceleration amplitude. \textbf{(b)} The phase difference, $\phi(t) = \theta_2(t)-\theta_1(t)$ of the spinners, of side length $L = 5.7$ mm, is found by tracking $\theta_1(t)$ and $\theta_2(t)$, comparing the orientation of the spinner at time $t$ (white) to a reference configuration (gray). \textbf{(c)} Oblique perspective of a pair of co-chiral spinners on the surface of a vibrating fluid bath. The self-generated capillary wavefield of the spinner is visualized by the distortions of a reflected color pattern \cite{harris2017visualization}. The shown experiment has parameters $\gamma/g = 1.6$, $f = 90$ Hz, and $d= 28$ mm with a wavelength of $\lambda = 3.56$ mm. The scale bar denotes $5$ mm.}
    \label{fig:ExpSetup}
\end{figure*}

With the universality of synchronization in coupled oscillators transcending both animate and inanimate systems, it is perhaps of no surprise that a shared {\it fluid} medium, wherein hydrodynamic forces and torques dictate interactions, enables synchronization behaviors at both the micro- \cite{tuatulea2022elastohydrodynamic, um2020phase,sabrina2018shape, zhang2023synchronization, riedl2023synchronization, aubret2018targeted, maggi2015micromotors, di2012hydrodynamic, niedermayer2008synchronization, golestanian2011hydrodynamic} and macro-scales \cite{alben2009wake,becker2015hydrodynamic,saenz2021emergent, oza2019lattices}. Of those, a subset focus on the synchronization of microscopic gears often referred to as spinners \cite{aubret2018targeted, maggi2015micromotors, di2012hydrodynamic, sabrina2018shape}. Here, micrometer-sized spinners are able to communicate via the surrounding fluid at low Reynolds' number ($\text{Re} = UL/\nu$). This regime is known as the Stokes' regime, wherein fluid-solid interactions are principally dominated by viscous stresses, with negligible influence of fluid inertia. In general, significantly less is known about propulsion and interactions at intermediate Reynolds numbers, where inertial stresses are comparable to those arising from viscosity \cite{klotsa2019above}. 

In recent years, the vibrating liquid interface has been used as a tunable and accessible tabletop experimental platform to probe fundamental questions in physics, often by analogy, ranging from topics in quantum mechanics, condensed matter physics, to active matter. Both droplets \cite{couder2005walking, protiere2006particle, pucci2015faraday, bush2020hydrodynamic} and floating solid objects \cite{ho2023capillary, barotta2023bidirectional, roh2019honeybees, rhee2022surferbot, thomson2023nonequilibrium, benham2024wave} have been shown to be able to harness energy from mechanical vibrations to produce directed motion via their self-generated wavefield. Multiple interfacial propulsors interact at long-range via the deformed fluid substrate \cite{couchman2022stability, thomson2020collectivedroplets, saenz2021emergent, nachbin2020kuramoto, ho2023capillary}. Phase synchronization has recently been documented for self-propelling droplets orbiting within small circular corrals that can communicate hydrodynamically via a thin intermediary fluid layer, giving rise to global order in large lattices \cite{saenz2021emergent}. The rich emergent interaction behaviors observed in these interfacial systems have been captured by mathematical modeling derived from first principles \cite{oza2013trajectory, oza2023theoretical, saenz2021emergent}, rather than needing to resort to ad-hoc phenomenological modeling, as is often common for companion dry granular systems where local steric interactions drive both propulsion and interaction.

We herein study the wave-mediated interactions of chiral solid objects atop a vibrating liquid interface.  Prior work documented that such capillary spinners steadily rotate via their self-generated wavefield when in isolation \cite{barotta2023bidirectional}.  When two spinners are held at a \textit{fixed} distance from one another, but otherwise left to rotate freely, we here find that the objects may still spin yet robustly synchronize their phase, with the equilibrium phase angle between the spinners depending sensitively on their spacing and vibration parameters.  In particular, the nature of the spinner-spinner coupling is shown to be spatially periodic, a direct consequence of their wave-mediated hydrodynamic interaction.  Spinners with different natural rotation rates can either synchronize or exhibit phase drift, depending on the degree of their intrinsic differences and the strength of the coupling.  Furthermore, when two spinners are placed sufficiently close to one another, their rotation can be completely arrested via their overwhelming mutual wave interaction. Our principal experimental observations are reproduced by a mathematical model that extends prior work on interacting capillary surfers (polar wave-propelled particles) \cite{oza2023theoretical} to the present case of interacting spinners.

\section*{Synchronization of Identical Spinners}

\begin{figure*}
    \centering
    \includegraphics{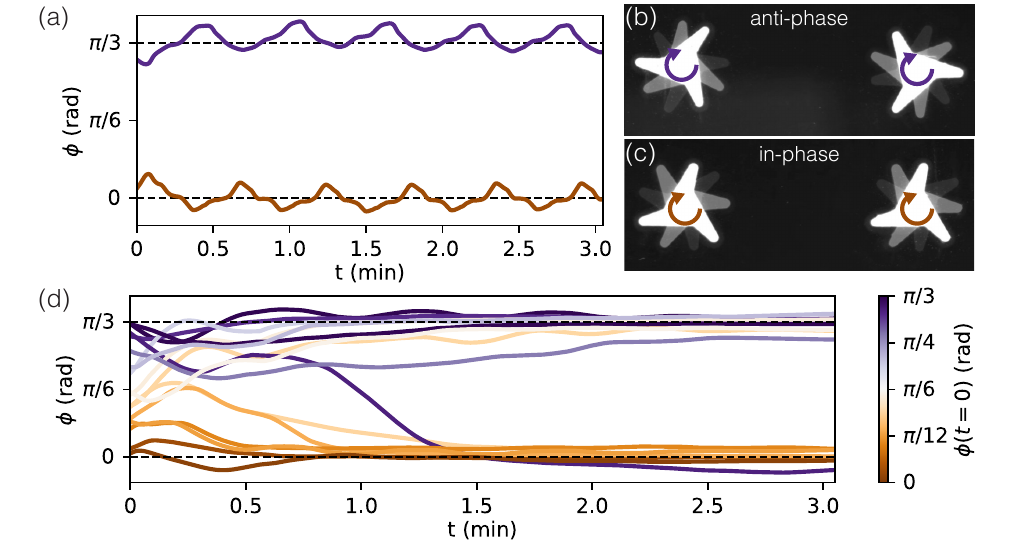}
    \caption{ \textbf{Synchronization of a pair of identical spinners via their mutual wavefield.} \textbf{(a)} Time-evolution plots of the phase difference ($\phi = \theta_2-\theta_1$) between two spinners demonstrates the bistability of in-phase and anti-phase modes for a fixed set of experimental parameters ($f = 90$ Hz, $\gamma/g = 1.6$, $d= 28$ mm). While the phase difference exhibits oscillations, the mean phase difference approaches either $0$ (in-phase) or $\pi/3$ (anti-phase) for different initial conditions. \textbf{(b-c)} Temporal evolution of the anti-phase and in-phase modes, corresponding to the purple and orange curves of \textbf{(a)}. \textbf{(d)} Initializing a pair of co-chiral spinners at different initial phase differences ($\phi(t=0)$) leads to either in-phase or anti-phase mode selection over an entire range of initial conditions. \edit{The color of each curve corresponds to the value of the initial condition}. The curves in \textbf{(d)} are smoothed over the period of one revolution to highlight the mean phase behavior for 16 independent trials. See Supplementary Video 1 for video of both in-phase and anti-phase synchronization.}
    \label{fig:Bistability}
\end{figure*}

A pair of spinners, of side length $L=5.7$ mm, are placed on a liquid (water-glycerol mixture) bath that is vertically vibrated with an acceleration amplitude $\gamma$ and angular frequency $\omega = 2\pi f$. The contact line is pinned along the base of the spinner, with its weight supported by buoyancy and surface tension. In order to restrict translational motion, we embed each spinner with a vertically polarized magnet at its center of mass and place a second magnet beneath the bath. This method allows for the center-to-center distance $d$ of a spinner pair to be prescribed for each experiment, allowing for a controlled investigation of the role of spacing on the wave-mediated interaction behavior (Fig. \ref{fig:ExpSetup}(a)). To assess the coordination of the two spinners, we measure their instantaneous phase difference $\phi(t) = \theta_2(t) - \theta_1(t)$ over time, with $(\theta_1(t),\theta_2(t))$ being tracked by comparing the current configuration of the spinner at a time $t$ to a reference configuration of the spinner (Fig. \ref{fig:ExpSetup}(b)). Extended details on the experimental setup and details on the tracking and post-processing can be found in Appendices A and B, respectively. 

The pair of spinners emit outwardly propagating capillary waves of wavelength $\lambda$ due to the relative motion of the spinner and the liquid bath (Fig. 1(c)). Emission of outwardly propagating capillary surface waves has been shown to be a means of propulsion both for translation \cite{roh2019honeybees, ho2023capillary, rhee2022surferbot} and rotation \cite{barotta2023bidirectional} along an interface. The excess momentum flux associated with the surface waves yields a net force/torque for translation/rotation whose magnitude can be tuned via the amplitude of the waves \cite{longuet1964radiation}. \edit{For the case of chiral spinners, it has been shown that surface wave-particle interactions can lead to propulsion \cite{barotta2023bidirectional, francois2018rectification, francois2020nonequilibrium}. In the present system, the waves (and any associated flows) originate from the driven vertical dynamics of the spinner \-- in the absence of the spinner, the fluid is at rest.  The excess momentum flux from wave generation drives steady rotation, with an angular velocity set by the balance of wave-mediated torques and viscous drag \cite{barotta2023bidirectional}.  In related works, chiral particles subjected to externally generated Faraday waves and concomitant flows can also rotate on the interface, with propulsion dependent on the geometry \cite{francois2018rectification, francois2020nonequilibrium}. In the present work, we restrict the study to driving amplitudes below the Faraday threshold wherein propulsion is driven strictly by the self-generation of waves by the spinner.}

When the three-armed spinner considered here is shaken at driving parameters $\gamma/g = 1.6$ and $f = 90$ Hz \edit{($\lambda = 3.56$ mm)}, a stable rotation of a constant angular velocity $\Omega = 0.19 \pm 0.03$ rad/s is observed.  This value is determined by taking the average and standard deviation of the mean rotation speed of 10 different spinners.  While this uncertainty reflects variations due to slight manufacturing differences between individuals, the speed of an single spinner exhibits much less variability over time, typically an order of magnitude less than the variation between individuals. When a second spinner is placed in the bath at a center-to-center distance $d = 28$ mm, the two spinners continue to spin by the same mechanism, but now also respond to their neighbor's incident wavefield.  

Two example trials with different initial conditions are shown in Fig. \ref{fig:Bistability}(a), which displays the evolution of their phase difference over time (Supplementary Video 1). After a brief transient, the spinners rapidly find a constant mean phase angle, $\overline{\phi}$, indicative of synchronization.  Oscillations about the mean phase angle can be observed, and are due to both the wave-mediated interactions and extraneous magnetic effects (additional discussion can be found in the Supplemental Material \cite{supp}). For this particular spacing and set of driving parameters, the spinners exhibit both in-phase and anti-phase modes with roughly equal probability. The anti-phase Fig. \ref{fig:Bistability}(b) and in-phase Fig. \ref{fig:Bistability}(c) synchronization appear as $\phi \approx \pi/3$ and $\phi \approx 0$, respectively, given the three-fold symmetry of the spinner. 

To more robustly survey the possible equilibrium behaviors at these parameters, a series of 16 independent trials are conducted with different pairs of co-chiral spinners (Fig. \ref{fig:Bistability}(d)), \edit{with each line corresponding to a pair and the color corresponding to the initial phase difference}.  Following a transient period, the spinners find either the in-phase or anti-phase mode with roughly equal probabilities.  In this plot, the phase difference is smoothed over the rotation period of the spinner to emphasize the mean phase behavior, and highlight the clustering near phase angles of $0$ (in-phase) and $\pi/3$ (anti-phase). Although only three minutes of observation are shown here in order to clearly show the transient period, stable synchronization has been observed for longer trials (see Supplemental Material\cite{supp}). Furthermore, if the system is externally perturbed, the spinners will rapidly return to either the in-phase or anti-phase mode, indicative of the system's bistability for these parameters.

\begin{figure}
    \centering
    \includegraphics{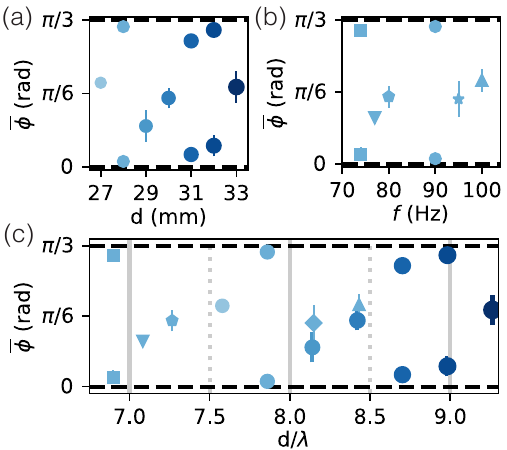}
    \caption{\textbf{Periodic behavior in the equilibrium mean phase difference ($\overline{\phi}$) for a pair of co-chiral spinners.} \textbf{(a)} Mean phase difference as a function of spinner spacing $d$, showing alternating behaviors of bi-stable in/anti-phase and intermediate phase difference. The frequency is fixed at $f=90$ Hz and acceleration amplitude at  $\gamma/g= 1.6$. \edit{Each shade of blue corresponds to a different distance $d$.} \textbf{(b)} Mean phase difference as a function of frequency $f$, showing alternating behaviors of bi-stable in/anti-phase and intermediate phase difference. The spacing is fixed at $d=28$ mm and amplitude of oscillation $a=\gamma/\omega^2=0.049$ mm. \edit{Each symbol corresponds to a different frequency $f$.}
    \textbf{(c)} For both cases, the mean phase difference is plotted as a function of the non-dimensional spinner spacing $d/\lambda$, demonstrating the overall periodicity of the phase behavior on the capillary wavelength. Error bars denote one standard deviation in the equilibrium mean phase difference.}
    \label{fig:DlambSweep}
\end{figure}

To understand whether this behavior is more generic, experiments are carried out over various values of driving frequency $f$, acceleration amplitude $\gamma$, and spinner spacing $d$ (Fig. \ref{fig:DlambSweep}). We first consider the role of the distance between the spinners.  Below the bath, the spacing of the tethering magnets can be adjusted in $1$ mm increments, ranging from $d=27-33$ mm (see Appendix A, Figure \ref{fig: ExplodedView} for more detail). For each spacing at least 6 trials are conducted. Further trials are conducted for distances admitting bistability of the in-phase and anti-phase modes. The phase difference is tracked over time, and the mean phase difference, $\overline{\phi}$, is recorded by smoothing $\phi$ over a time window equal to one period of rotation and extracting the ending value at the end of the trial. For all trials used, it was ensured that the mean phase angle had become constant. By fixing the driving parameters ($f=90$ Hz, $\gamma/g = 1.6$), it can be seen that as the distance increases, the mean phase behavior, $\overline{\phi}$, follows a non-monotonic trend with alternating regions of in-phase and anti-phase bistability (as in Fig. 2) interrupted by spacings that admit only intermediate phase angles, clustered roughly around $\overline{\phi} \approx \pi/6$ (Fig. 3(a)). \edit{We note that such intermediate phase angles are where the pair finds an equilibrium phase difference closely centered to that value and does not represent an intermittent switching between the in-phase and anti-phase modes defined above.} Mean phase angle for independent trials at a fixed distance were divided into two groups using a k-nearest neighbors (KNN) algorithm. If the difference in the two groups means was greater than $\pi/12$, we classified the system to be in a bistable regime, and reported the results as two separate populations.

Next, the frequency of the bath's vibration is varied while the spacing is held fixed at $d=28$ mm (Fig. \ref{fig:DlambSweep}(b)). \edit{The frequency is varied between $74$ Hz and $100$ Hz corresponding to wavelengths ranging from $\lambda = 4.21$ mm to $3.32$ mm, respectively.}  To ensure comparisons could easily be made between experiments at variable frequency, we fix the amplitude of oscillation $a = \gamma / (2\pi f)^2 = 0.049$ mm between experiments to approximately keep the wave amplitude constant between trials. Similar to the distance sweep, as the frequency of oscillation varies, the mean phase difference follows a similar alternating pattern, with the bi-stable in-phase/anti-phase behavior being observed at $74$ Hz in addition to the $90$ Hz case previously showcased in Fig. \ref{fig:Bistability}(a,d). 

To gain insight as to why similar alternating trends may be arising in both the distance and frequency sweeps, we consider the phase behavior as a function of a dimensionless distance between the two spinners, normalizing the spacing $d$ by the capillary wavelength $\lambda$.  Using the dispersion relation for capillary waves \cite{lamb1924hydrodynamics, de2004capillarity}, we can relate to the wavelength $\lambda$ to the experimental parameters via the capillary wave dispersion relation $\lambda =  \left(2\pi\sigma / \rho f^2\right)^{\frac{1}{3}}$. By plotting our data from Fig. \ref{fig:DlambSweep}(a) and Fig. \ref{fig:DlambSweep}(b) together with this new abscissa, regions of alternating phase behavior align, and are separated by approximately one wavelength.  This periodic interaction behavior directly reflects the wave nature of the hydrodynamic coupling and motivates the development of a dynamical wave-interaction model in what follows.

\begin{figure}
    \centering
    \includegraphics{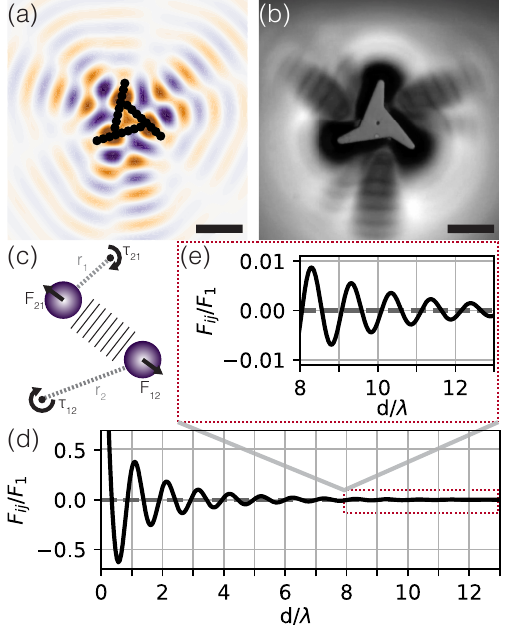}
    \caption{ \textbf{Simulating the wave interactions of capillary spinners.} \textbf{(a)} To simulate the wavefield of the spinner, a collection of 24 point wave sources are used to trace out the approximate spinner geometry which recovers the qualitative shape of the experimental wavefield at $f=90$ Hz. \textbf{(b)} Experimental spatial structure of the wavefield can be visualized via a shadowgraph technique shown here at $f= 90$ Hz and $\gamma/g = 1.6$ demonstrating qualitative agreement with $\textbf{(a)}$. Scales bars for $\textbf{(a,b)}$ denote $5$ mm. See Supplementary Video 2 for video of the isolated spinner self-propelling due to the emitted capillary waves. \textbf{(c)} Each point source acts as both an emitter of waves and also as a location wherein forces, and hence torques, can act from the other spinner. $\textbf{(d)}$ Two wave-emitting point sources exert forces on one another that lead to a non-monotonic, sign-alternating, force profile as a function of their effective distance $d/\lambda$ (Eq. \ref{eqn:force}). \textbf{(e)} Notably, the force at large effective distances remains oscillatory and finite, allowing for a weak coupling to persist between the spinners. The weak coupling strength ensures that the spinners are still able to effectively rotate under their own self-propulsion for most parameter regimes and large effective distances.}
    \label{fig: SimSetup}
\end{figure}

\section*{A model of synchronization}
In order to further interrogate and isolate the mechanism for synchronization, we develop a simple model of the wave-mediated interactions between two spinners that rotate about fixed centers (Fig. \ref{fig: SimSetup}). Each spinner is modeled as a collection of wave-emitting point sources ($J=24$ point sources per spinner) that allow us to approximately resolve their relatively complex shape (Fig. \ref{fig: SimSetup}(a)). As the point source spacing is significantly smaller than the wavelength ($\sim \lambda/5$), introducing additional sources has little effect on the predicted wavefield but increases the computational time of the model. The solution for the wavefield associated with a single oscillatory point forcing on a semi-infinite fluid bath of density $\rho$, viscosity $\nu$, and surface tension $\sigma$ is known from prior work \cite{de2018capillary, oza2023theoretical}.  The governing equations for the weakly viscous response to a localized parametric point forcing with angular frequency $\omega=2\pi f$ are given by potential flow in the bulk
\begin{align}
\nabla_{||}^2\Phi + \Phi_{zz} &= 0,
\end{align}
coupled to linearized dynamic and kinematic boundary conditions at the free surface ($z=0$) 
\begin{align}
-\Phi_t + \frac{\sigma}{\rho}\nabla_{||}^2h + 2\nu\nabla_{||}^2\Phi + gh+  \frac{F_0}{\rho}\cos\omega t\delta(\mathbf{x}) &= 0, \\
     \Phi_z + 2\nu \nabla_{||}^2h - h_t &= 0,
\end{align}
where $\Phi$ is the velocity potential, $h$ is the wave height, and $\nabla_{||} = \partial_{xx}+\partial_{yy}$.

The strength of the forcing is given by $F_0 = \alpha m \gamma$, where $\alpha=0.15$ is an experimentally determined parameter computed by matching the observed wave amplitude for a single isolated spinner ($\gamma/g = 1.6$, $f=90$ Hz) to that predicted by the model.  Future models could include an additional coupled equation for the vertical response of the floating solid body, as in \cite{benham2024wave}, in order to predict this parameter from first principles in the absence of experimental measurements. A more in-depth discussion on $\alpha$ and the manner in which it is determined is provided in Appendix B. 

The governing equations \edit{have previously been} solved for the wave-profile of a single source $h(r,t)$ \cite{oza2023theoretical}.  By expressing the height profile as $h(r,t) = \mathbf{Re}\left[H(r)e^{i\omega t}\right]$, the complex amplitude of the wavefield $H(r)$ \edit{is given as}
\begin{equation}
H(r) = \frac{F_0}{4\sigma} \sum_{p=1}^4 \frac{H_0(-k_pk_cr) - Y_0(-k_pk_cr)}{3+4i\epsilon/k_p + 4\epsilon^2k_p+\beta/k_p^2},
\end{equation}
where $H_n$ and $Y_n$ are the $n^{th}$ order Struve function and Bessel function of the second kind, respectively, and the $k_p$s are the roots to the polynomial $p(k;\epsilon,\beta) = \epsilon^2k^4 + 2ik^2 + k^3 +\beta k - 1$ as a function of the dimensionless parameters $\epsilon$ and $\beta$. The four parameters appearing in the result for the wavefield profile are 
\begin{align}
    k_c & = \left(\frac{\rho \omega^2}{\sigma}\right)^{1/3}, &  \quad \epsilon & = \frac{2\nu k_c^2}{\omega}, \nonumber \\
    \quad l_c & = \sqrt{\frac{\sigma}{\rho g}}, \quad & \beta & = \frac{1}{(k_cl_c)^2},\nonumber
    \end{align}
and correspond to the wavenumber ($k_c$), a reciprocal Reynolds number ($\epsilon$), the capillary length ($l_c$), and a wave Bond number ($\beta$).  This wavefield has been predicted in prior work, and successfully applied to model the interactions of capillary surfers \cite{oza2023theoretical}. The wavefields of each individual point source comprising the spinner are superposed to predict the overall shape of the wavefield (Fig. \ref{fig: SimSetup}(a)).  The predicted wavefield of an individual spinner shows good qualitative agreement with the equivalent experimental visualization (Fig. \ref{fig: SimSetup}(b)) (Supplementary Video 2), in particular, capturing the pronounced constructive wave ``beams'' emerging from each wedge.  In other systems, complex geometric designs of microgears have prompted similar modeling using a discrete distribution of point sources \cite{di2012hydrodynamic}.  For each spinner, the constituent point sources remain geometrically constrained relative to each another, however the overall structure of 24 sources may rotate about its center of mass in response to various self-induced and external torques described next.

To model the interactions of two spinners, a second spinner is translationally fixed at a center-to-center distance $d$ from the first.  Mathematically, we can then describe the system by two second-order ODEs, one for each spinner's phase $\theta_n(t)$ (with $n=1,2$).  Each spinner is assumed to be driven by a self-propulsion and wave-interaction term, yielding the governing equations
\begin{align}
 I\ddot{\theta}_n = -\frac{I}{{\tau}_{\nu}}\left(\dot{\theta}_n-\Omega_n\right) + \sum_{i=1}^{J}\sum_{j=1}^{J} \left[ \mathbf{r}_{i} \times  \mathbf{F}_{ij}\right] \cdot \hat{\mathbf{z}},
 \label{eq:ode}
\end{align}
where $J$ is the number of point sources per spinner $(J=24)$, $I$ is the moment of inertia, ${\tau}_{\nu}$ is the viscous timescale, $\Omega_n$ is the speed of an isolated spinner, $\mathbf{r}_i$ is the position of point source $i$ relative to the center of the spinner, and $\mathbf{F}_{ij}$ is the wave-force between point sources, computed next. \newedit{The moment of inertia is computed in Autodesk Fusion, a Computer-Aided Design (CAD) program, for the exact geometry and material used in experiment. This does not, however, include possible added mass effects associated with the acceleration of fluid around the body.}  

The time-averaged force between two point particles can be found via the equation for the height of the wavefield as in prior work \cite{de2018capillary} (Fig. \ref{fig: SimSetup}(c-e)). We assume that wave force on particle $i$ from the wavefield of particle $j$ is given by $\mathbf{F}_{ij}=\left\langle\left. m_i \ddot{z}_i \nabla h\left(\mathbf{x}-\mathbf{x}_j, t\right)\right|_{\mathbf{x}=\mathbf{x}_i}\right\rangle$ where $\langle \cdot \rangle$ represents the time average over one oscillation period and $\ddot{z}_i = -\alpha \gamma \cos(\omega t)$.
Assuming that the two particles oscillate on the fluid interface with the same amplitude and phase, this time-averaged force is given by,
\begin{equation}
        \mathbf{F}_{ij}(r) = F_1\sum_{p=1}^4 \text{Re}\left[ 
 \frac{H_{-1}(-k_pk_cr)+Y_{1}(-k_pk_cr)}{1+(4/3)i\epsilon/k_p + (4/3)\epsilon^2k_p+\beta/3k_p^2}\right]\hat{\mathbf{r}}_{ij},
 \label{eqn:force}
\end{equation}
where $F_1 = \alpha^2 m_im_j\gamma^2k_c/24\sigma $ and $\hat{\mathbf{r}}_{ij} = \frac{\mathbf{x}_j-\mathbf{x}_i}{|\mathbf{x}_j-\mathbf{x}_i|}$.

\begin{figure}
    \centering
    \includegraphics{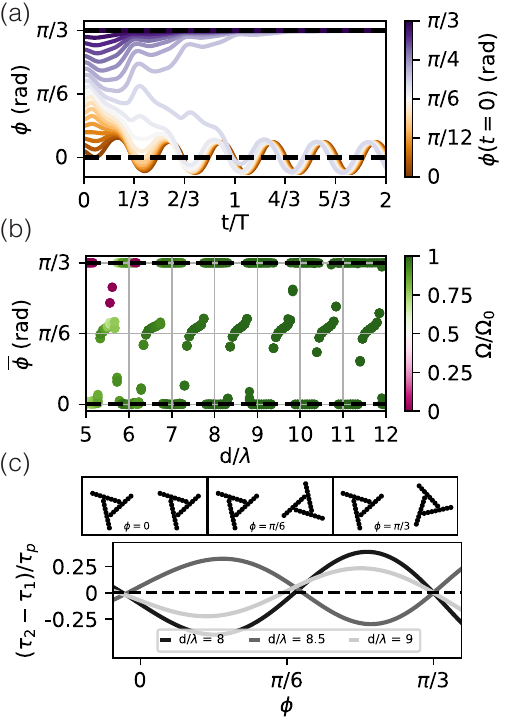}
    \caption{ \textbf{Simulated wave-mediated interactions of a pair of spinners.} \textbf{(a)} Like experiment, bi-stability in the mean phase behavior is recovered for parameters corresponding to the experiments in Fig. \ref{fig:Bistability}. \textbf{(b)} The dependence of the mean phase angle as a function of effective distance is recovered. Notably, bistable regimes are separated by intermediate zones with phase clustering at $\sim \pi/6$. In addition, a phase locking behavior with the spinners experiencing cessation of rotation is predicted for small effective distances (top left corner in burgundy) wherein the rotation speed $\Omega$ goes to zero following an initial transient. \edit{\textbf{(c)} The difference in wave interaction torques $\tau_2-\tau_1$ between the two spinners as a function of their relative phase difference. Here, $\theta_1=\pi/6$ and $\theta_2$ is varied. We find that the location and stability of the fixed points (zeros) vary as a function of the effective distance, $d/\lambda$, elucidating the seen trends in the equilibrium phase difference in experiment and simulation.}}
    \label{fig: SimRes}
\end{figure}

The two ODEs ($n=1,2$) are coupled via the wave-interaction term. In the absence of this term, the spinner finds an equilibrium steady rotation at speed $\Omega_n$, which we extract from experiments on an isolated spinner ($\Omega_n = 0.19$ rad/s at $90$ Hz and $\gamma/g = 1.6$).  This value, along with the viscous timescale $(\tau_{\nu}$ = 0.53 s), and the waveheight scaling factor ($\alpha=0.15$) are parameters in our model obtained from experiments on a single spinner. The viscous timescale is found by starting a spinner from constant rotation at $\Omega_0$ and shutting off the vibration leading to a slowdown response that is fit to an equation of the form $\Omega(t) = \Omega_0e^{-t/\tau_{\nu}}$. Further details on the viscous timescale measurement can be found in Appendix B. Given the relatively large distances $d \gg l_c$ between spinners, we omit the capillary attraction forces in the present work, as $F_c \sim e^{-d/l_c}$ \cite{ho2019direct} and $d\gg l_c\approx 2.4$ mm.

The two-spinner model is solved numerically using parameters corresponding to the experiments shown in Fig. 3.  Notably, the simulation is able to recover the observed bi-stability (Fig. 5(a)), with the mean phase difference approaching either 0 (in-phase) or $\pi/3$ (anti-phase) after a brief transient. \newedit{For both experiment and simulation, we find that a brief transient persists before reaching the equilibrium mean value. The transient period persists for roughly 1-2 rotations of the spinner for the parameters explored herein, with the duration of the transient generally increasing with spinner spacing.} \edit{Scenarios with an initial phase difference close to 0 tend to recover the in-phase mode whereas initial phase differences closer to $\pi/3$ tend to recover the anti-phase mode. For all of the simulations shown, the initial velocities are set to be equal to the spinner's angular velocity in isolation ($\Omega_n$). It is difficult to similarly prescribe consistent initial velocities across all experimental trials, and thus may explain why there is less direct correlation between the initial phase difference and the outcome (in-phase or anti-phase synchronization) for the corresponding experimental data in Fig. 2(d).}
   We additionally explore the role of spinner spacing $d$ in the simulation, with the results of the mean phase summarized in Fig. 5(b). 

Analogous to the experimental findings in Fig. \ref{fig:DlambSweep}, a distinct periodicity is observed in the mean phase behavior, a direct consequence of the periodicity in the force between two point sources over the characteristic length $\lambda$. For each $d/\lambda$ tested, eight initial conditions $\phi(t=0)$ are chosen for the spinner pairs (uniformly sampled between 0 and $\pi/3$) to ensure that all possible modes of interaction are uncovered.

\edit{To gain further insight as to why the spinners select either a bistable state, with both in-phase/anti-phase synchronization coexisting, or an intermediate phase angle, we appeal to a reduced analysis of the wave interaction torques in the problem (Fig. 5(c)). Here, we compute the difference of wave-mediated torques between a spinner pair, $\tau_2-\tau_1$, as a function of the phase difference, $\phi$. \newedit{We normalize the interaction torques from waves by the self-propulsion torque $\tau_p =I\Omega_n / \tau_{\nu}$}.  This torque difference arises when subtracting the individual evolution equations for $\theta_1$ and $\theta_2$, and thus suggests an evolution for the phase difference $\phi$. The self-propulsion torques cancel for identical spinners, and if one neglects inertia, the phase difference evolution is strictly governed by the difference of the two wave interaction torques (i.e. $\dot{\phi}\sim\tau_2-\tau_1$).  For the plot in Fig. 5(c)), we fix $\theta_1=\pi/6$ and vary the value of $\theta_2$, as depicted in the graphics next to the plot.  We find that when bistability is the preferred state (at integer multiples of the wavelength, e.g. $d/\lambda = 8,9$), stable fixed points occur at $\phi\approx 0, \pi/3$, with an unstable fixed point near $\phi\approx\pi/6$. In cases when an intermediate phase difference is preferred, we see that a single stable fixed point occurs at a value near $\pi/6$, with the points near $0$ and $\pi/3$ becoming unstable. While highly simplified, this analysis represents a first step towards rationalizing this wave-mediated synchronization phenomena.}

While the interaction has a clear periodicity, it also has a spatial decay (via both radial spreading and viscous attenuation), that leads to additional trends in the predicted mean rotation speeds of the spinner. For relatively distant spacing, the spinners rotate at a mean speed very close to their isolated free speed, but still manage to synchronize.  Their mean rotation speed is reduced as they move closer, with the wave coupling becoming stronger.  The simulations further allude to the possibility that in some cases the wave coupling may become sufficiently strong to result in a total stifling of rotation, reminiscent of the ``amplitude death'' phenomenon for highly coupled phase oscillators \cite{mirollo1990amplitude, yamaguchi1984theory}.  In this regime, the spinners synchronize their phase to a fixed value but cease to rotate entirely, suggesting that it is possible for the wave-interaction torque to exceed that from self-propulsion.  We demonstrate that this behavior is also readily observable in experiment later. A more expansive discussion on the simulation is found in Supplemental Material highlight the robustness of the main phenomena, and trends associated with changes in various model parameters\cite{supp}.

\section*{Synchronization of Non-Identical Spinners}

\begin{figure*}
    \centering
    \includegraphics{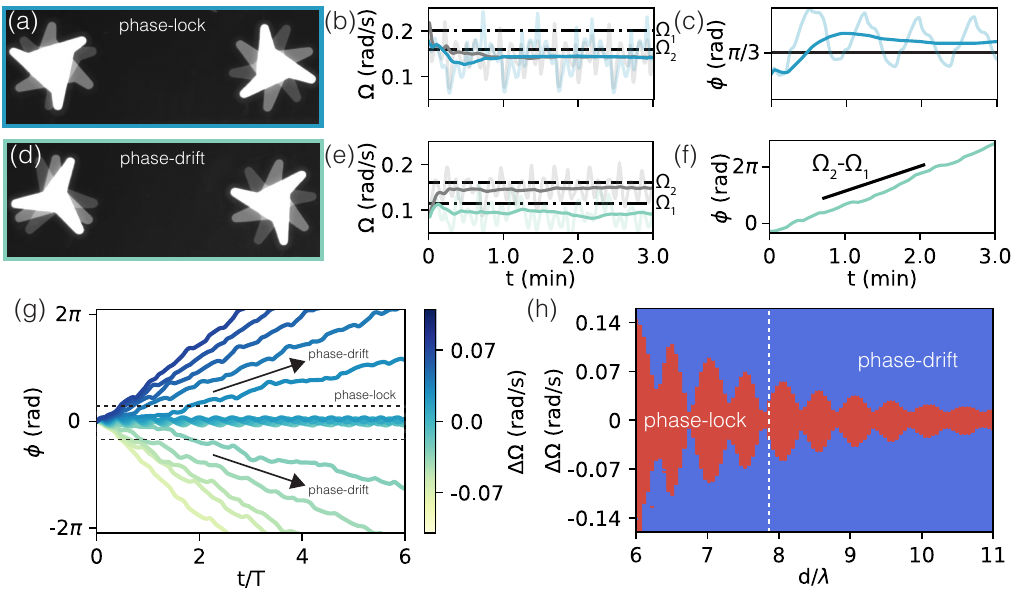}
    \caption{\textbf{Synchronization of non-identical spinners.} With slight modifications to the spinner geometry, the intrinsic angular velocity can be tuned in experiment. \textbf{(a)} When the difference in free speed between two non-identical spinners is small, the pair can synchronize and phase lock. \textbf{(b)} The spinners are able to synchronize. Here, a pair of spinners with free angular velocity $\Omega_1 = 0.20 \pm 0.01$ rad/s and $\Omega_2 = 0.16 \pm 0.003$ rad/s exhibit anti-phase synchronization \textbf{(c)}, with $\overline{\phi}\approx \pi/3$. The smoothed phase difference over one period of rotation is shown opaquely, with the raw data faintly presented in the background. \textbf{(d)} When the difference in angular velocity is too large, the pair instead experiences phase drift and are unable to synchronize. \textbf{(e)} Here, a pair of spinners of free angular velocity $\Omega_1 = 0.114 \pm 0.016$ rad/s and $\Omega_2 = 0.16 \pm 0.003$ rad/s cannot compromise their intrinsic difference in angular velocity. All experiments are conducted at $f=90$ Hz, $\gamma/g = 1.6$, and $d = 28$ mm. \textbf{(f)} The phase difference grows at a relatively constant rate roughly equal to the difference in angular velocities, $\Omega_2-\Omega_1$, indicating that synchronization of the two spinners is not achieved. In \textbf{(b,e)}, the gray and colored line denote the angular velocity of the two spinners smoothed over a period of oscillation, with corresponding raw data more faintly in the background. Dashed and dash-dot lines indicate free (isolated) speed for the two spinners. \textbf{(g)} Simulations are able to recover both phase-drift and phase-locking as a function of $\Delta \Omega = \Omega_2-\Omega_1$ within a small region of $\Delta \Omega = 0.02$ rad/s allowing for synchronization to occur at $d/\lambda = 7.9$. $\Omega_1$ is held fixed at $0.19$ rad/s for all simulations. \textbf{(h)} An expansive sweep over distance $(d/\lambda)$ and $\Delta \Omega$ demonstrate that the admissible $\Delta \Omega$ is a function of the effective distance $(d/\lambda)$ following a functional form reminiscent of the magnitude of the force between two wave-emitting point sources. The dashed white line indicates the value of $d/\lambda$ used in \textbf{(g)}. See Supplementary Video 3 for video of both phase-locking and phase-drift for a pair of non-identical spinners.}
    \label{fig: Kuramoto}
\end{figure*}

In the study of phase oscillators, a natural question to consider is whether synchronization remains possible when there exists intrinsic differences between the individuals. Perhaps the most well-known model of such a system is the Kuramoto model \cite{kuramoto1975self, kuramoto1984chemical,acebron2005kuramoto}. In this model system, if the difference in intrinsic or natural angular frequency of a pair of oscillators is small, synchronization can still occur, indicated by the phase difference $\phi$ ``locking'' at a particular value.  For larger differences in their intrinsic frequencies, the differences are not able to be reconciled and instead the phase difference ``drifts''; increasing or decreasing at a rate approximately equal to the differences in their two natural frequencies $\dot{\phi} \approx \Omega_2-\Omega_1$.

Although in a real experiment no two spinners are truly identical, to the best of our ability we have demonstrated that synchronization is a robust outcome for two ''identical'' spinners over a range of distances and frequencies.  This result is similarly recovered in our simulations with identical spinners \edit{but can also be explored for spinners with intrinsic differences}.  The angular velocity of a spinner in isolation depends sensitively on its geometry \cite{barotta2023bidirectional}. We here modify the chirality of spinner design to achieve other angular speeds. While holding the overall size of the spinner and wedge opening angle $(\psi = 7\pi/9)$ constant, the ratio of the long to short side in each wedge is altered. For our original spinner design, the short to long wedge arm length ratio is given by $q=0.6$. We can continuously vary this ratio, leading to spinners with different intrinsic angular velocities (Fig. \ref{fig: Kuramoto}). The limiting cases of $q=0$ and $1$ correspond to achiral spinners.  At $\gamma / g = 1.6$, $f= 90$ Hz, and $d = 28$ mm, we establish a proof of concept for synchronization between non-identical spinners. 

When the difference in angular velocity is small (Fig. \ref{fig: Kuramoto}(a-c)) $(\Omega_1,\Omega_2) = (0.20\pm 0.01,0.16 \pm 0.003)$ rad/s, phase locking is possible, and a mode reminiscent of the anti-phase mode is recovered. Here the original design of short-to-long wedge arm ratio, $q = 0.6$, is altered to $q=0.25$ resulting in the speed change. The mean phase difference is not exactly $\pi/3$, presumably due to the difference in natural frequencies. When the difference in angular velocity is larger (Fig. \ref{fig: Kuramoto}(d-f) $(\Omega_1,\Omega_2) = (0.114 \pm 0.016,0.16 \pm 0.003)$ rad/s, phase drift instead occurs. Here, the disparate spinners are unable to compromise their speeds to achieve a synchronized motion brought about by modulating the short to long wedge arm ratio of $0.76$. The time rate of change of the phase difference is roughly constant at a value near the difference in their intrinsic angular velocities ($\overline{\dot{\phi}} \approx 0.05$ rad/s), with fluctuations in the speed and phase difference due to the complex wavefield (Supplementary Video 3). This demonstrates that not only does this system offer a unique platform to study synchronization over an identical population, but it is easily tunable via geometric design to tackle more general notions of synchrony in polydisperse populations. 

While our experiments for non-identical pairs offer a proof-of-concept motivating future experimental studies, our simulation allows for a more finely controlled parametric exploration of this behavior.  For the purposes of this numerical study, all parameters in the model are fixed to the values corresponding to $\gamma/g = 1.6$, $f=90$ Hz, and $d=28$ mm, however the value of the free speed $\Omega_1$ is varied with $\Delta\Omega=\Omega_2-\Omega_1$.  In the experimental realization using different spinner geometries, a number of parameters are changing in practice, making a direct comparison more cumbersome.  In Fig. \ref{fig: Kuramoto}(g) we show that the simulation is able to recover cases where synchronization is and is not possible depending on the magnitude of $\Delta\Omega$.  For these simulations, we initialize all spinner pairs with a phase angle of $\phi(0) = 0$ and initial velocities equal to their free speeds in isolation.  In particular, we find that there is a band of $\Delta\Omega$ centered around 0 where synchronization (phase locking) is still achieved. In contrast, when $|\Delta \Omega|$ becomes too large, phase drift occurs instead, indicated by the phase difference increasing at a roughly constant rate.

In addition, the breadth of the synchronous band of $\Delta \Omega$ varies as a function of the effective distance (Fig. \ref{fig: Kuramoto}(h)). As the effective distance grows larger, the coupling strength generally is reduced leading to a smaller range of $\Delta \Omega$ admitting synchronization.  However, there is also a distinct periodicity in the extent of these regions. The curve separating the phase locking and phase drifting behaviors follows a functional form similar to that of the elemental wave interaction force $|F_{ij}|$ between two point sources. \edit{In particular, the spatial decay of the envelope and the periodicity of the wave interaction force are reflected in the admissible region of phase-locking.} With these various observations as inspiration, future work could attempt to deduce a generalized Kuramoto (GK) model, similar to what is derived in \cite{saenz2021emergent} for other wave-mediated synchronizing systems.

\section*{High Coupling and Amplitude Death}

\begin{figure*}
    \centering
    \includegraphics{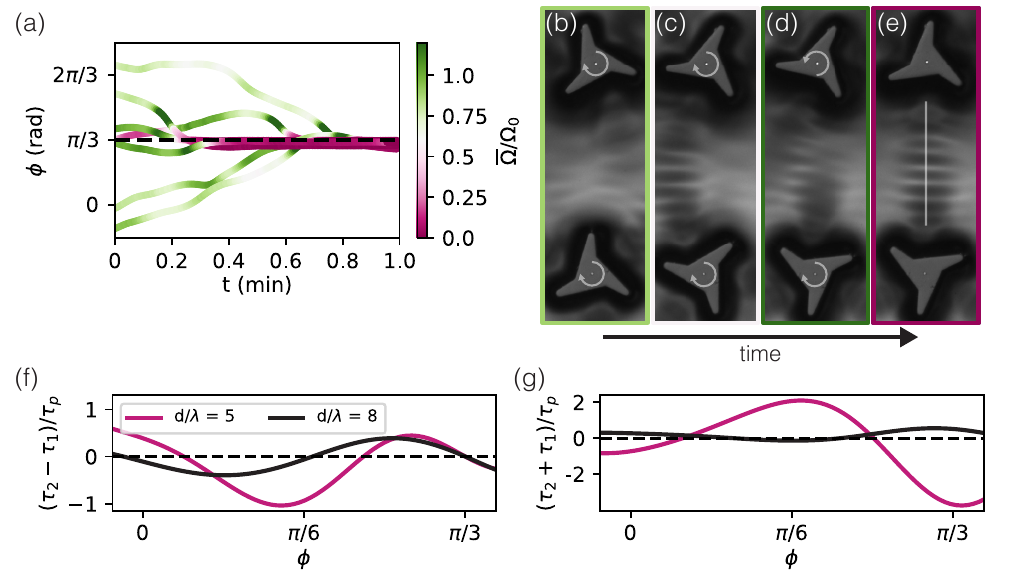}
    \caption{\textbf{Amplitude death of two spinners in the large coupling limit.} \textbf{(a)} When a pair of co-chiral spinners are highly coupled (corresponding to a small effective spacing $d/\lambda$ and large amplitude waves), the amplitude death (zero angular velocity) state of the spinners is observed. Here, the pair of spinners find a preferred orientation causing the pair to stop spinning at $d= 28$ mm at $70$ Hz $(d/\lambda = 6.6)$. After a crescendo in mean angular speed, the locking state is found, and the pair lock into the stable state shown by the lighter color in plot wherein $\overline{\Omega} = (|\Omega_1|+|\Omega_2|)/2$. \textbf{(b)-(e)} is a time sequence demonstrating the route to locking. Notably, in \textbf{(d)}, the pair overshoot the equilibrium configuration causing a reversal in the direction of the top spinner before settling into the locking state ($\phi \approx \pi/3$) in \textbf{(e)}. \textbf{(b)-(e)} is color-coded to align with the dynamic rearrangement process as shown in \textbf{(a)} over time. See Supplementary Video 4 for a video of an amplitude death event for a pair of spinners. \newedit{ \textbf{(f)} The predicted difference of torques is shown as a function of $\phi$, holding $\theta_1=\pi/6$ and varying $\theta_2$ using the same parameters as in Fig. 5(c). For both, $d/\lambda = 5$ and $d/\lambda=8$, stable equilibrium phase differences exist at $\phi \approx 0,\pi/3$. \textbf{(g)} When considering the sum of torques, it is clear that the phase difference at $\pi/3$ has a different signature for the two different effective distances. Notably, at $d/\lambda = 5$, the sum of torques is largely negative allowing for the amplitude death state to be reached by balancing self-propulsion. }}
    \label{fig:AmpDeath}
\end{figure*}

For the experimental results presented thus far, {\it dynamic} modes of synchronization are achieved over a range of spacing and frequencies, with a mean phase angle between the spinner pair depending sensitively on the effective distance between the two objects. In this regime, the wave-mediated force is sufficiently weak as to not overwhelm the self-propulsion behavior, but still allow for meaningful communication.  As suggested by our simulations, we herein demonstrate that further strengthening the coupling can allow for {\it static} modes where the resulting speed of each spinner is zero, and the phase is locked indefinitely.

The effective coupling strength can be enhanced by decreasing the relative spacing between the spinners, either by decreasing their dimensional spacing or by decreasing the driving frequency. For pendula, parameter combination that are generally deemed ``high coupling'' are able to produce amplitude death (alternatively named ``beating death'') \cite{bennett2002huygens, zou2021quenching} featuring the complete cessation of oscillation.  In our system, this can be interpreted as the halting of spinner rotation. For the parameters studied in Fig. \ref{fig:DlambSweep}, the coexistence of dynamic in-phase and anti-phase was demonstrated.  However if the frequency is lowered from 90 to 70 Hz (holding the driving amplitude $\gamma/\omega^2$ constant) while keeping the distance fixed at $d=28$ mm ($d/\lambda = 6.6$), the two spinners cease rotation, and lock each other in place (Fig. \ref{fig:AmpDeath}(a)). The state is stable for all accelerations tested $\gamma/g = 0.9 - 1.25$, with higher acceleration exploration limited by the Faraday wave threshold $(\gamma_F/g \approx 1.3$). We expect that amplitude death is also possible at $90$ Hz, as alluded to by simulation, although in experiment this would require going to smaller effective distances wherein extraneous magnetic interactions between the spinners would become non-negligible.

For all initial conditions tested, the spinners reorient and eventually collapse into a stable static configuration with a phase difference of $\approx \pi/3$. The transient dynamics generally feature an overall slowdown Fig. \ref{fig:AmpDeath}(b-c) of the spinner rotation due to the strong wavefield interaction. However, immediately before the complete cessation of rotation, a large crescendo in the speed is observed Fig. \ref{fig:AmpDeath}(d), where the speed of an individual spinner can momentarily exceed its free speed. Thereafter, the spinners remain locked in place with a standing wave pattern visible between them Fig. \ref{fig:AmpDeath}(e). In this regime, coupling via the wavefield is strong enough to fully overcome the intrinsic self-propulsion (Supplementary Video 4). 

In some cases, the spinner pair may even overshoot their desired equilibrium phase angle of $\pi/3$ and then reverse rotation direction before reaching the static state.\newedit{ The amplitude death state can also be interpreted in terms of the simplified analysis introduced in Fig. 5(c) for cases where the anti-phase mode is a stable fixed point (Fig. \ref{fig:AmpDeath}(f-g)), wherein $\dot{\phi}\approx 0$ or $\dot{\theta_1}\approx\dot{\theta_1}$. In addition to the difference of torques $\tau_2-\tau_1$ being approximately zero (Fig. \ref{fig:AmpDeath}(f)), the \textit{sum} of the two torques in the anti-phase configuration is largely negative (Fig. \ref{fig:AmpDeath}(g)). Again neglecting inertia, the sum of the two equations of motion suggests a simplified model $\dot{\theta_1} + \dot{\theta_2} \sim 2\tau_p + \tau_1+\tau_2$. As such, if the sum of $\tau_1+\tau_2$ is negative and of sufficient magnitude relative to $\tau_p$, the wave-mediated torques may fully offset the self-propulsion torque yielding the static state.  To illustrate this, we compare the prior case of dynamic synchronization occurring at $d/\lambda = 8$ to an amplitude death state in simulation at $d/\lambda = 5$ using the same parameters as in Fig. 5(c). Notably, while both cases admit equilibrium phase angle solutions of $\overline{\phi} \approx 0,\pi/3$, the sum of the wave torques ($\tau_1+\tau_2$) is a large negative value near $\pi/3$ for $d/\lambda = 5$. }

\section*{Discussion and Outlook}

We have presented a fluid-mediated system of capillary spinners that self-propel and interact via their mutually generated wavefield. For this study, we have fixed the distance between the objects in order to focus solely on the synchronization of their rotational dynamics. It was found that the equilibrium phase difference between a spinner pair can be tuned as a function of the effective distance, $d/\lambda$, and the system exhibits bi-stable regimes of both in-phase and anti-phase synchronization for a fixed set of experimental parameters. While dynamic regimes were the principal focus of the current study, it was also found that static modes are present with the pair ceasing rotation and locking into a configuration with a fixed phase angle. Finally, experimental tuning of the spinner geometry leads to variations in their intrinsic rotation speeds (analogous to natural frequencies in other oscillator systems) allowing for the system to also be readily applied to study interactions of non-identical oscillators.  In general, the interaction behaviors uncovered in this study show distinct signatures of the underlying wave-based coupling.

Given the simplicity and tunability of the experimental setup, questions centered on globally coupled oscillator networks could be considered. In the simplest extension of the current setup, one could untether the spinners by removing the embedded magnets to consider the case where oscillators can both sync and swarm. It has been shown that when analogous polar objects called capillary ``surfers'' are placed on a vibrating liquid bath, the surfers exhibit spontaneous positional order and swarming behaviors with a variety of stable modes even for the case of a single pair \cite{ho2023capillary, oza2023theoretical}. In this study, we have demonstrated that surface waves can also lead to rotational synchronization as well which sets up a natural followup of the possibility of simultaneous syncing and swarming. Such oscillators that sync and swarm, known as ``swarmalators,'' have been a focus of many recent theoretical and numerical studies \cite{o2017oscillators, o2019review, o2022collective} with experimental realizations slowly beginning to emerge \cite{barcis2019robots}. The setup presented here is an ideal candidate given that the spinners would be all pairwise coupled to one another via the fluid interface, with the additional benefit that the governing equations of motion are known and amenable to stability analysis \cite{oza2023theoretical}. We expect that for an untethered spinner pair, the spinners would synchronize their phase by the wave torque mechanism elucidated here, but also adapt their center-to-center spacing to satisfy the lateral force balance acting on the spinners.

With such a setup in place, moving to large collections of untethered, or tethered, spinners could also be used to study emergent phenomena. On much smaller lengthscales, chiral collections have recently gained attention in both living \cite{tan2022odd} and artificial \cite{bililign2022motile, soni2019odd, bricard2013emergence, bricard2015emergent} systems with emergent phenomena such as symmetry reversals \cite{workamp2018symmetry}, large boundary flows in confinement \cite{yang2020robust}, lane formation \cite{reeves2021emergence}, and self-reverting vortices \cite{caprini2024self} being realized. In bouncing droplet systems, which use a similar experimental setup to the one used here, moving to higher collections of interacting drops has already demonstrated various dynamic modes such as soliton formation \cite{thomson2020collectivedroplets}, magnetic order \cite{saenz2021emergent}, and collective motion in droplet rings \cite{couchman2020free}, behaviors that depend sensitively on the lattice structure and number of droplets used. In general, most of the emergent behaviors documented in the bouncing droplet system are isolated to narrow regions of parameter space due to the increased subtleties associated with the droplet's vertical dynamics and stringent criteria for maintaining non-coalescence behavior. We envision similar types of global phenomena to emerge in the wave-mediated spinner and surfer systems, with the potential for additional effects due to their customizable geometry and their robust engineerable propulsion characteristics.  

Such spinner systems have the intrinsic characteristic that their geometry features a form of $n-$fold symmetry. As such, one would expect that the pairing or bonding between a spinner pair to be dictated by their symmetries. Large lattice formation is thus anticipated to be dictated by the shape of the constituent units (spinners) and could result in geometric frustration if a prescribed lattice arrangement is incompatible with the shape of the spinners themselves and their preferred pairwise interaction modes \cite{han2008geometric, mellado2012macroscopic, ortiz2019colloquium}.

Furthermore, active solids have been a recent topic of interest \cite{baconnier2022selective, fersula2024self} with applications to odd elasticity, for instance. Flexible, yet steric, connections between active agents exhibit a zoo of behaviors, such as when confined to various lattice structures \cite{baconnier2022selective, baconnier2024self}. Rather than having rigid connections like the aforementioned active solid, the contactless, yet interacting, spinners show promise for being a system where fundamental questions related to odd viscosity might be probed \cite{fruchart2023odd}. Given the chiral design of the particle and transverse interactions between adjacent spinners in the plane, the system holds promise to probe further questions in chiral active fluids with a simple tabletop setup \cite{khain2023trading, banerjee2017odd, soni2019odd}.

\section*{Acknowledgements}

The authors would like to acknowledge Nilgun Sungar and Tali Khain for stimulating discussions. The authors gratefully acknowledge the financial support of the National Science Foundation (NSF CBET-2338320). J.-W.B. is supported by the Department of Defense through the National Defense Science and Engineering Graduate (NDSEG) Fellowship Program. G.P. acknowledges the CNR-STM Program. A.H. acknowledges the Hibbitt Fellowship.

\section*{Author contributions}
J.-W.B., G.P., and D.M.H. designed research;
J.-W.B., G.P., and A.H. conducted preliminary experiments; J.-W.B. and E.S. designed final experimental setup; J.-W.B. developed and performed simulations and final experiments; J.-W.B. and D.M.H. analyzed data, developed models, and wrote the paper; all authors performed research, discussed the results, commented on the manuscript and gave final approval for publication, agreeing to each be held accountable for the work performed therein; D.M.H. supervised the project and secured funding.

\section*{Appendix A: Experimental Methods}

\subsection*{Fabrication of the spinners}
 
The method for fabricating the spinners follows prior work \cite{barotta2023bidirectional}.  The spinners are fabricated using OOMOO 30, a commercially available two-part (Part A and B) silicone rubber compound.  Once well-mixed, the mixture is poured into 3D-printed (Formlabs Form 3+, Clear Resin V4) molds designed in Autodesk Fusion 360. A small central extrusion is added to each spinner mold to create a cylindrical pocket for the magnet to be inserted.  The CAD file associated with the molds used in the present work are provided with other data presented throughout at the associated digital repository: \href{https://github.com/harrislab-brown/SyncSpinners}{https://github.com/harrislab-brown/SyncSpinners}. Rounded corners at the three tips of the spinner were added to help maintain the pinned contact line along the bottom perimeter throughout the course of the experiments. To pop larger bubbles that formed during the mixing process, and to allow the mixture to make its way into some of the smaller crevices in the mold, the mixture was poured from a height of approximately 1 meter. Excess OOMOO was then scraped off the top of the mold using a straight-edge razor blade. The molds containing the mixture were placed in a vacuum chamber for a total of at least 120 seconds in four 30 second intervals to remove any smaller air bubbles present in the mixture. The surface of the molds were then scraped firmly one final time leaving little-to-no excess compound. The liquid mixture was then cured at room temperature for at least 6 hours, typically overnight. Once cured, the 2 mm-thick spinners were removed from the mold via compressed air.  The spinners are then embedded with cylindrical Neodymium 52 magnets (Supermagnetman-Cyl0003-50) of $0.5$ mm height and diameter $0.3$ mm. The cured silicone-rubber compound is naturally hydrophobic. When placed on a water-glycerol bath, the spinners are supported by surface tension and hydrostatic pressure, and remain level with the contact line pinned along the perimeter of their bottom face.   

 \subsection*{Experimental setup and protocol}
  A circular fluid bath (depth $H=5.7$ mm and diameter $120$ mm) is mounted atop a rigid aluminum mounting platform and driven vertically using an electrodynamic modal shaker (Modal Shop, Model 2025E). The mounting platform is connected to the shaft of a linear air bearing that interfaces with the shaker via a long flexible stinger, ensuring uniaxial motion at the platform \cite{harris2015generating}.  For all frequencies and amplitudes explored in the present work, the vertical vibration amplitude was uniform to within 1\%. The acceleration amplitude was monitored using two accelerometers (PCB Piezotronics, Model 352C65) mounted diametrically opposed on the driving platform, and a control loop maintained a constant driving amplitude.  The entire vibration assembly is mounted on a vibration isolation table (ThorLabs ScienceDesk SDA75120) to minimize any influence of ambient vibrations.

\begin{figure}
    \centering
    \includegraphics{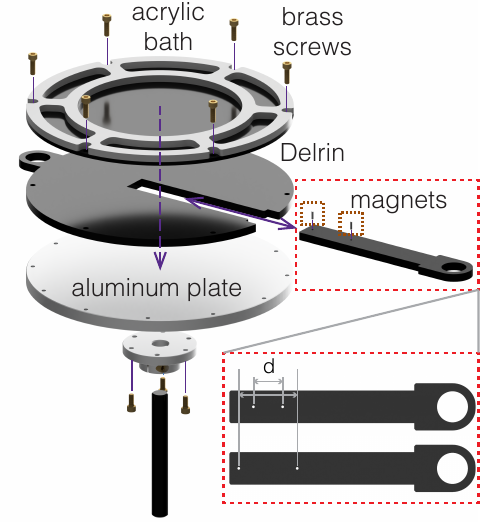}
    \caption{ \textbf{Expanded view of the experimental setup.} Exploded view of experimental equipment design for magnetic pinning of interfacial spinners. The top plate is made of two layers of laser cut PMMA (Acrylic) bonded with a Methylene Chloride based solvent (Weld-On 4). The central $120$ mm diameter circular bath is filled with a water-glycerol mixture during experiments. The middle plate is made of laser cut acetal resin (Delrin) with removable inserts for variable magnet spacing (corresponding to spinner spacing $d$). The bottom machined aluminum plate mounts to a vertical shaft constrained by a linear air bearing and is actuated by a modal shaker.}
    \label{fig: ExplodedView}
\end{figure}

  Below the bath, cylindrical Neodymium 42 magnets (K\&J Magnetics-D14) of $6.25$ mm height and diameter $1.58$ mm are placed in laser-cut Delrin sheets allowing for the vertical alignment of the magnet-laden spinners (see Fig. \ref{fig: ExplodedView}). This effectively locks the spinners in place with minimal lateral motion of the center of mass of the floating objects and no static friction.
 
 The fluid used in all experiments was a water-glycerol mixture with a density of $\rho = 1149.5 \pm 0.5$ kg m$^{-3}$ measured daily with a density meter (Anton Paar DMA 35A).  The water-glycerol mixture has a surface tension of $\sigma = 0.067 \pm 0.001$ N m$^{-1}$ \cite{glycerine1963physical} and a dynamic viscosity of $\mu = 0.01 \pm 0.0005$ kg m$^{-1}$ s$^{-1}$ \cite{cheng2008formula}. 

 For each experiment, $64 \pm 1$ mL of the water-glycerol solution was measured using a graduated cylinder and poured into the circular bath constructed from laser-cut sheets of acrylic. The working fluid depth ($5.7$ mm) coincides with the thickness of the acrylic sheet to avoid the formation of a border meniscus known to generate small capillary waves that radiate into the bath even in the absence of the spinners \cite{shao2021role}. The fluid in the bath was changed at least every thirty minutes to minimize contamination effects. A ring of LEDs was placed $3.8$ cm above the bath to illuminate the spinners (appearing white) against the fluid (appearing black).

Once the bath was filled, each spinner was thoroughly rinsed with deionized water and dried with a dry cleaning wipe (Kimtech Science) before being placed on the bath with a pair of clean tweezers. In the absence of vertical vibration of the bath, the spinner and fluid remained stationary. If the spinner was placed on the bath, and the contact line had any visible imperfections, the spinner was promptly removed and the process restarted.

\section*{Appendix B: Data Acquisition, Processing, and Parameter Measurement}

 \subsection*{Tracking rotational motion of the spinners}
 
 The spinners are filmed for data acquisition at 3 frames-per-seconds (fps) using an overheard camera (Allied Vision, Mako U-130B) with a macro lens placed 1 meter above and normal to the bath surface. Images were processed in MATLAB: the video frames are run through the geometric transformation estimator $\texttt{imregtform}$ with \texttt{tformType} set to \texttt{rigid}, thus determining both translation and rotation ($\Delta \theta$) of the spinner as compared to a reference image at $t=0$. To obtain the initial phase difference between the spinner pair, the algorithm was additionally run, comparing the two initial frames. Shadowgraph videos were taken at higher frame rates for visual clarity. 

 \subsection*{Determination of the mean phase angle}

The smoothed curves of the phase difference $\phi$ were computed using Matlab's $\texttt{SmoothMean}$ routine over a trailing number of frames equal to the amount of frames in one rotation of the spinner. For example, if the period of revolution for the spinner was $30$ seconds, the smoothing routine would be done over a trailing window of $90$ frames (corresponding to the 3 frames-per-second recording).  The smoothed phase angle $\overline{\phi}$ denotes the mean phase angle over the last period of revolution in the experimental recording or simulations.  All trials had reached a steady state for at least one full rotation.

\subsection*{Determination of model parameters}

When running simulations of the spinner pair dynamics, experimental parameters for the spinner speed in isolation, viscous timescale, and wave amplitude prefactor are needed. This section describes how those parameters are determined from experiments on single spinners.

The spinner speed in isolation is simply the constant angular velocity achieved by a spinner for a fixed set of driving parameters i.e. $0.19 \pm 0.03$ rad/s at $\gamma/g = 1.6$ and $f= 90$ Hz. A value of $\Omega_n=0.19$ rad/s is used in the simulations, unless otherwise stated.

\begin{figure}
    \centering
    \includegraphics{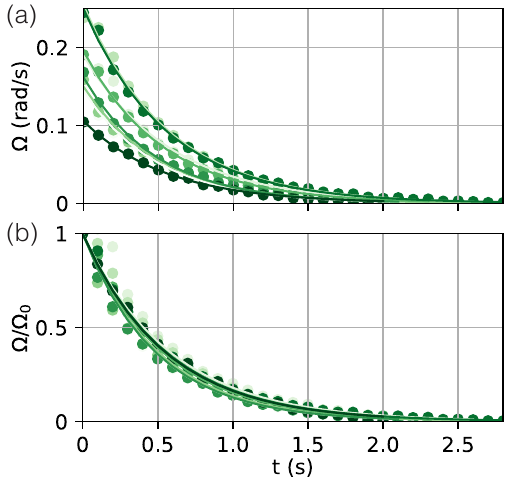}
    \caption{\textbf{Viscous timescale measurement.} \textbf{(a)} A spinner is rotating at a constant speed $\Omega_0 = \Omega(t=0^-)$ when the acceleration amplitude is reduced to zero $(\gamma / g = 0)$ at $t=0^+$ causing the speed to exponentially decay to zero. Markers shown denote experimental measurements of the angular velocity over time, and the best-fit curves for $\Omega(t) = \Omega_0e^{-t/\tau_{\nu}}$ are shown for fitting constants $\Omega_0$ and $\tau_{\nu}$. \textbf{(b)} By rescaling the $y$ axis by $\Omega_0$, the time constant is evidently consistent across trials, taking a value of $\tau_{\nu} = 0.53 \pm 0.03$ s.  }
    \label{fig: FigViscous}
\end{figure}

The viscous timescale can also be measured with a single spinner on the bath Fig. \ref{fig: FigViscous}. Noting that the proposed model for the spinner assumes a \textit{linear} (viscous) drag, it then follows that when the self-propulsion torque is removed (by dropping the vibration amplitude of the bath to zero) the angular velocity of the response should follow a decaying exponential of the form $\Omega(t) = \Omega_0e^{-t/\tau_{\nu}}$, where $\Omega_0$ and $\tau_{\nu}$ are the angular velocity right before shutting off the vibration amplitude and and viscous timescale, respectively. 

By starting the spinner at different initial angular velocities (achieved by altering the acceleration amplitude at a fixed value of frequency of $f= 90$ Hz) 8 trials were conducted and then the measured speed fit to a decaying exponential of the form $\Omega(t) = \Omega_0e^{-t/\tau_{\nu}}$ (Fig.  \ref{fig: FigViscous}(a)), with both $\Omega_0$ and $\tau_{\nu}$ being determined by the fit. The resulting experimental measurement for the viscous timescale is $0.53 \pm 0.03$ s (Fig. \ref{fig: FigViscous}(b)) across all trials. A value of $\tau_{\nu}=0.53$ s is used for all simulations.  We expect the viscous timescale to depend on both geometric design of the spinner, properties of the liquid used, and depth of the bath. We do not expect the viscous timescale to depend on the driving parameters.

In order to determine a value for the wave amplitude prefactor $\alpha$, the wavefield of a single spinner was measured experimentally using the Fast Checkerboard Demodulation (FCD) method \cite{wildeman2018real} for 90 frames (corresponding to $2.02$ s).  The modulus of the wave height at each pixel location was then time-averaged over all frames.  A single amplitude value was then determined by averaging the mean wavefield over a small annular region surrounding the spinner ($10\leq R \leq 10.2$ mm). The corresponding wave model for a single spinner was then simulated over one period of oscillation and then averaged in time and over the same annular region as in the experiment. The value of $\alpha$ was then set so that the average value in the annulus was the same between experiment and simulation. The final result of $\alpha = 0.15$ was robust, as it was recovered for a range of annuli radii and widths tested.  If the annular radius was too large, the experimental measurements are noisy (due to the small wave height) leading to inconsistent results.  Similarly if the annular radius was too small, large distortions due to the static meniscus and spurious processing defects associated with the region near the solid body, led to reduced measurement accuracy.

\begin{figure}
     \centering
     \includegraphics{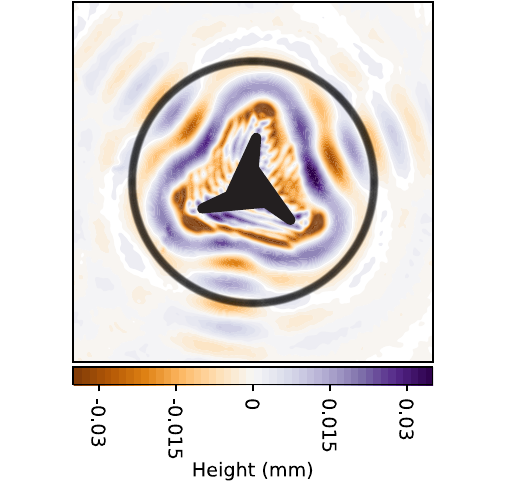}
     \caption{\textbf{Example wave amplitude measurement.} Measured spinner wavefield for a driving parameters of $\gamma/g = 1.6$ and $f=90$ Hz. Wave amplitudes are on the order of 10 $\mu$m.  The thin \edit{black} ring indicates the annular region over which the wave amplitude was averaged in time and space to determine the wave amplitude prefactor $\alpha$ used in simulation.   }
     \label{fig: WaveMeasure}
 \end{figure}

\section*{Data availability}

The final experimental data shown for generating Fig. \ref{fig:Bistability},\ref{fig:DlambSweep},\ref{fig: Kuramoto},\ref{fig:AmpDeath} in the study and CAD files for the spinner molds for the shapes used is available at \href{https://github.com/harrislab-brown/SyncSpinners}{https://github.com/harrislab-brown/SyncSpinners}.

\section*{Code availability}

Code used for simulating a spinner pair is available at \href{https://github.com/harrislab-brown/SyncSpinners}{https://github.com/harrislab-brown/SyncSpinners}.

\section*{Competing interests}
The authors declare no competing interests.

\section*{Additional information}
\textbf{Supplementary information} 4 Supplementary Videos and Supplemental Material are available.

\bibliography{refs}

\end{document}